# A sparse spin qubit array with integrated control electronics

J. M. Boter[1,2,*], J. P. Dehollain[1,2,3,*], J. P. G. van Dijk[1,2], T. Hensgens[1,2], R. Versluis[1,4], J. S. Clarke[5], M. Veldhorst[1,2], F. Sebastiano[1], and L. M. K. Vandersypen[1,2,5,†]

[1]QuTech, Delft University of Technology, The Netherlands, [2] Kavli Institute of Nanoscience, Delft University of Technology, The Netherlands, [3]School of Mathematical and Physical Sciences, University of Technology Sydney, Australia [4]Netherlands Organisation for Applied Scientific Research (TNO), The Netherlands, [5]Components Research, Intel Corporation, USA

[*]These authors contributed equally to this work [†]Mail: Lorentzweg 1, 2628 CJ Delft, The Netherlands, l.m.k.vandersypen@tudelft.nl

*Abstract*—Current implementations of quantum computers suffer from large numbers of control lines per qubit, becoming unmanageable with system scale up. Here, we discuss a sparse spin-qubit architecture featuring integrated control electronics significantly reducing the off-chip wire count. This quantum-classical hardware integration closes the feasibility gap towards a CMOS quantum computer.

## I. INTRODUCTION

Semiconductor spin qubits [1, 2] are an attractive platform for large-scale quantum computers, due to their potential compatibility with well-established semiconductor manufacturing processes. In the last decade we have witnessed tremendous progress in the development of spin-qubit hardware [3-8] and significant interest and contribution of the semiconductor industry into spin-qubit research [9-11]. Therefore, the open questions surrounding the challenge of scaling-up [12] have become timely and highly relevant. One of the main issues in common with all nanoelectronic qubits is that current implementations require at least one external control line for every qubit. The small pitch of quantum dots (Fig. 1) permits extremely dense qubit arrays but aggravates the interconnect challenges. Existing proposals for dense 2D spin qubit arrays [13, 14] assume either a device density or material homogeneity that remains to be achieved. Another approach involves a network architecture, where qubits are arranged in small-cluster nodes, interconnected by long-range entanglement distribution channels, with the goal of creating space for easing the density requirements of the interconnects [12]. The feasibility of implementing quantum error correction protocols using this approach has been thoroughly analyzed [15], but the description of the physical implementation is largely missing. Here, we present a design of a sparse two-dimensional array whereby classical electronics integrated locally with the quantum hardware is used to minimize the need for off-chip interconnects and hence with a scalable Rent's exponent [16]. We first describe the components of the array and the implementation of quantum gates and measurements, followed by a description of the control electronics required to operate the qubits in the array and correct errors via the surface code approach [17]. We then analyze how this implementation of locally integrated electronics reduces the number of connections at the quantum plane boundary, and the required footprint of such components. Finally we provide a discussion of some of the technological considerations, potential challenges and options for solving them.

## II. ARRAY DESIGN

We propose a quantum computing architecture consisting of a two-dimensional array of electron-spin qubits using linear arrays of gate electrodes (Fig. 1) arranged to form a square lattice of electrostatically defined quantum dots with nearest-neighbor connectivity. In conventional spin qubit designs every quantum dot, with a typical pitch smaller than 100 nm, hosts a qubit. The proposed design uses a *sparse* qubit array with the qubits separated by ~12 μm and the vertices connected via electron shuttling channels to transport electrons to and from interaction regions. The array's sparseness enables the integration of sample-and-hold circuits alongside the quantum dot circuitry allowing to offset the inhomogeneities in the potential landscape across the array by independent DC biasing while sharing the majority of control signals for qubit operations across the array. The latter allows for a significant reduction in the number of connections at the quantum plane boundary. As detailed in Fig. 2, we start from a 22 mm × 33 mm (726 mm$^2$) die. The qubits are defined in the *quantum plane*, a section of the die consisting of $M \times M$ *modules*, each containing $N \times N$ *unit cells*. The unit cell is the smallest operational unit, containing four qubits along with all the elements required to operate them, as described in Fig. 3. Qubits remain at the vertices of the lattice while idle and are shuttled to the operation regions between the vertices in order to perform single- and two-qubit operations as well as readout and initialization.

## III. DC BIASING

Fig. 4 shows circuit schematics of locally integrated sample-and-hold circuits providing individual DC biasing of all the control gates, which total 64 gates per unit cell. Gate voltages within a unit cell are updated sequentially via four local demultiplexers that each distribute DC voltages generated remotely (i.e., outside the quantum plane) to 16 local capacitors connected to the gates. All demultiplexers within a module share the same input DC biasing signal, and all demultiplexers in the quantum plane share the same address bus (see Fig. 4(f)). The demultiplexers are enabled sequentially and in turn sequentially update each gate. In this way, all modules are updated in parallel and therefore one module refresh cycle is required to refresh the entire qubit array. We define two bias voltage resolutions, based on the gate functionalities. Gates acting as barriers to shuttling channels only require a resolution sufficient to maintain an electron in a quantum dot and therefore we can afford a coarse resolution of 1 mV. All other plunger and barrier gates require a resolution of 1 μV [12]. The minimum hold capacitance required to achieve the coarse resolution is ~0.16 fF (limited by the electron charge $e/\Delta V$), while the fine resolution requires ~14 pF (limited by thermal noise $k_B T/C$, assuming power dissipation from the local electronics raises the operating temperature to 1 K). The gate voltage refresh rate will be set by the current leakage of the hold circuit and the time required to update each gate, which in turn will set the module size (i.e., the number of unit cells, and therefore total gates, which can be sequentially updated).

## IV. SIGNALS FOR QUBIT OPERATIONS

All the qubit operations are performed by shuttling the qubits to the operation regions and applying pulsed signals to the appropriate gates to perform the operations.

The shuttle channels are defined by a linear array of gates (blue gates in Fig. 3), along which a traveling wave potential can trap and shuttle an electron. The traveling wave potential is generated using four phase-shifted sinusoidal signals on four consecutive gates (different shades of blue in Fig. 3), with the signals being reused every fifth gate. The shuttling signals are always on and the phase shifts control the direction of shuttling. With the use of a barrier gate (Fig. 3(b)), an electron can be forced to tunnel into a shuttle channel. The traveling wave potential is made large enough to overcome the inhomogeneities in the potential landscape, eliminating the need to apply DC biasing on the shuttling gates.

Single-qubit gates are performed by applying a microwave pulse to the control gate labelled MW in Fig. 3(c). A pair of micromagnets in this operation region provides magnetic field gradients required to perform electric dipole spin resonance (EDSR) [18]. A two-qubit gate is performed by pulsing the control gate labelled J, to activate an exchange interaction between electrons underneath the adjacent gates. In order to apply the AC signals on gates that also require DC biasing, we make use of a complementary switching circuit (see $\varphi_{AC}$ and $\overline{\varphi_{AC}}$ in Fig. 4(b)).

The surface code is sustained using a cyclic sequence of pulsed signals within a unit cell, with the same sequence performed in parallel across all unit cells in the entire array. A set of remote pulsed voltage sources is used to generate the required cyclic pulsed signals at each gate (i.e., one source per pulsed gate in a unit cell). Logic gates in the surface code with lattice surgery are achieved by creating defects in the lattice. We implement these defects by preventing shuttling of a subset of data qubits via locally integrated switches.

## V. READOUT

Qubit readout is performed at the operation region shown in Fig. 3(c). A readout quantum dot connected to source/drain ohmic contacts is used for charge sensing and spin readout is achieved via spin-to-charge conversion based on Pauli spin blockade [2]. Additionally, the ohmics in this region provide electrons that are shuttled to the unit cell vertices to initialize the array.

The drain contacts of all readout dots in a module are connected to a single line at the quantum plane boundary, and readout is performed sequentially across the unit cells of each module, while the modules are read out in parallel. This is achieved by sequentially pulsing every sensor plunger in a module to the low-impedance, electrostatically sensitive regime, while all other sensors in the module are at high-impedance (i.e., Coulomb blockade). The sequential control of the plungers in a module is achieved using a global readout demultiplexer that can be shared between all modules across the entire array.

## VI. LINE SCALING

The signal routing we have described (as summarized in Table 1), enabled by the described DC biasing scheme, allows for a very efficient scaling of the ratio of connections needed at the unit cell level to connection outputs at the quantum plane boundary. Considering that the total number of gates scales with the number of qubits ($4M^2N^2$), we now discuss how the operation schemes described above allow scaling down the number of connections at the quantum plane boundary.

The sparse array with sample-and-hold circuits provides independent DC biasing with $O(M^2+N)$ lines at the quantum plane boundary. All pulsed and microwave control signals needed to sustain the surface code, can be shared across every unit cell in the entire array. This amounts to a constant number of 58 lines at the quantum plane boundary irrespective of the number of qubits. The signals used to control the switches that deactivate data qubits for the logical qubit implementation scheme, are arranged in a crossbar fashion across the entire quantum plane, reducing the number of lines for this purpose to as few as $O(MN)$ at the quantum plane boundary. In practice, we propose to use $x$ crossbars over the entire array, in order to allow for $x$ defects to be simultaneously created and manipulated, bringing the line scaling to $O(xMN)$. By using decoding to address the readout plungers per module, the sequential readout scheme obtains a line scaling at the quantum plane boundary as $O(M^2+log(N))$. At the boundary of the quantum plane, Rent's exponent can thus be as low as $p=0.5$.

## VII. FOOTPRINT

We now consider the footprint requirements of the control electronics that need to be locally integrated in the quantum plane, and the wire density at various levels.

The bulk of the footprint will be taken up by the capacitors required for the sample-and-hold scheme. Coarse resolution is required for 32 gates and another 32 gates require fine resolution, which comprise a total capacitance per unit cell of ~450 pF. Assuming ~1 pF/μm$^2$ (using state-of-the-art deep-trench capacitor technology [19]), we estimate a total capacitor footprint of ~450 μm$^2$. In addition, we modelled a demultiplexer circuit using 40-nm technology, extrapolated to 28-nm technology and obtained an estimate of the total footprint of the DC biasing and readout demultiplexers of ~60 μm$^2$ per unit cell. This adds to a total footprint per unit cell of ~510 μm$^2$, which allows to set the qubit pitch to $d \geq 12$ μm. Assuming a 50 nm pitch between gate electrodes (Fig. 1), this would require linear arrays of 240 gate electrodes per lattice arm. A unit cell has an area $4d^2 \approx 576$ μm$^2$ and a perimeter $8d \approx 96$ μm. O($10^2$-$10^3$) wires pass through the unit cell perimeter, using multiple interconnect layers. The area and perimeter for a module are $(2dN)^2$ and $8dN$, respectively, and the quantum plane has an area and a perimeter of $(2dNM)^2$ and $8dNM$, respectively.

For a total of $2^{20}$ ($\approx 10^6$) qubits, the total area covered by the quantum plane is ~151 mm$^2$, leaving ~575 mm$^2$ of space remaining in the die, which can be used to implement classical control circuits and to bring the wire count going off-chip, typically the real bottleneck for Rent's rule, to well below the wire count at the quantum plane boundary by means of additional levels of multiplexing.

## VIII. DISCUSSION AND OUTLOOK

Very-large scale spin qubit devices will ultimately be based on a trade-off of a large number of considerations. With this proposal we explore the extreme sparse approach, with single qubits placed at the nodes of the shuttling channels. Different from some existing proposals, this approach does not make strong assumptions on the potential landscape homogeneity or the density with which transistors and qubits can be integrated, but it does assume that spins can be shuttled over 10 μm distances with very high fidelity. It should also be possible to design a similar integrated electronic scheme for architectures with larger qubit-cluster nodes, for which it has been shown that the fidelity requirements of the shuttling channels are more relaxed [15]. We also assume that magnetic field inhomogeneities and g-factor variations can be overcome by individual dc tuning.

| Shuttling gates (blue) | Source → gate |
|---|---|
| Pulsed gates (red) | DC: source → local demultiplexer → gate<br>AC: source → gate |
| Sensing dot plunger (purple) | DC: source → local demultiplexer → gate<br>AC: source → global demultiplexer → gate |
| Drain contacts | Measurement device ← ohmic |

Table 1. Summary of signal routing for the four different type of control lines in the array design.

In this work we have focused on the reduction of the number of control lines at the quantum plane boundary, as well as on the footprint of the classical control electronics, a key first step to assess the feasibility of implementing sparse spin qubit architectures, which motivates future work into addressing the following open issues. Distributing all signals for qubit operations across the entire qubit array requires careful design for minimizing crosstalk, along with to estimate the total line capacitance, which will affect clock speeds and required source power. There will be a large number of switches, used to separate the cycles of DC biasing and qubit operation on the applied gate voltages and to perform lattice surgery. The power dissipated by these switches can be significant and will be a factor in considering the clock rates of the system and the achievable operating temperature. It is most likely that both the surface code cycle rate and the size of the array will be limited by the number of sequential readouts required, since this is the most time consuming of all operations. Some degree of parallel readout can be implemented by amplitude modulation or frequency modulation. If that is not sufficient, smaller readout modules can be defined, each consisting of a subset of unit cells that are read sequentially. This comes at the expense of an increased number of readout connections.

All things considered, this proposal provides an appealing outlook for the long-term implementation of larger scale quantum computing chips, and provides guidance for near-term research at the quantum, classical and integrated levels.

## REFERENCES

[1] D. Loss, and D. P. DiVincenzo, *Phys. Rev. A* **57**, 120 (1998).
[2] F.A. Zwanenburg *et al.*, Rev. Mod. Phys. **85**, 961 (2013).
[3] E. Kawakami *et al.*, Nat. Nano **9**, 666 (2014).
[4] M. Veldhorst *et al.*, Nat. Nano **9**, 981 (2014).
[5] J. Yoneda *et al.*, Nat. Nano **13**, 102 (2018).
[6] T. F. Watson *et al.*, Nature **555**, 633 (2018).
[7] X. Xue *et al.*, Phys. Rev. X **9**, 021011 (2019).
[8] W. Huang *et al.*, Nature **569**, 532 (2019).
[9] R. Maurand *et al.*, Nat. Comm. **7**, 13575 (2016).
[10] P. Galy *et al.*, IEEE J. of the Electron Devices Society **6**, 594 (2018).
[11] R. Pillarisetty *et al.*, in 2018 IEEE Int. Elec. Dev. Meet. (IEDM) (IEEE, New York, 2018), pp. 6.3.1-6.3.4
[12] L. M. K. Vandersypen *et al.*, npj Quantum Inf. **3**, 34 (2017).
[13] M. Veldhorst *et al.*, Nat. Comm. **8**, 1766 (2017).
[14] R. Li *et al.*, Sci. Adv. **4**, eaar3960 (2018).
[15] B. Buonacorsi *et al.*, Quant. Sci. Tech. **4**, 025003 (2019).
[16] D. P. Franke *et al.*, Microprocessors and Microsystems **67**, 1 (2019).
[17] C. Horsman *et al.*, New J. of Phys. **14**, 123011 (2012).
[18] M. Pioro-Ladrière *et al.*, Appl. Phys. Lett. 90, 024105 (2007).
[19] J. M. Park *et al.*, in 2015 IEEE Int. Elec. Dev. Meet. (IEDM) (IEEE, New York, 2015), pp. 26.5.1-26.5.4

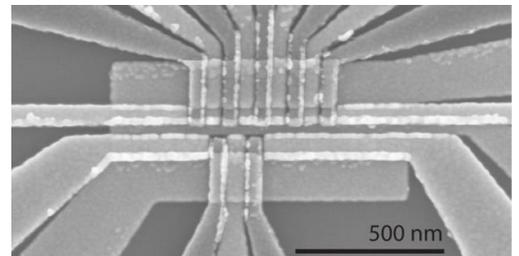

Fig. 1: Image of a set of gate electrodes from a state-of-the-art device of electrostatically defined quantum dots.

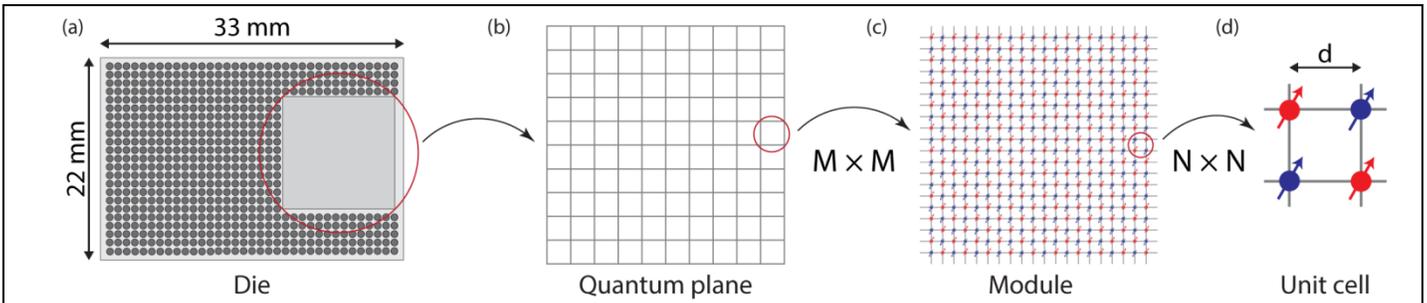

Fig. 2 Overview of the qubit architecture with a schematic breakdown of its components as described in the main text, including (a) the die containing (b) the quantum plane area, highlighting a single (c) module which contains a set of (d) unit cells. Qubits are color coded to distinguish data qubits (blue) and ancilla qubits (red), as defined in the surface code.

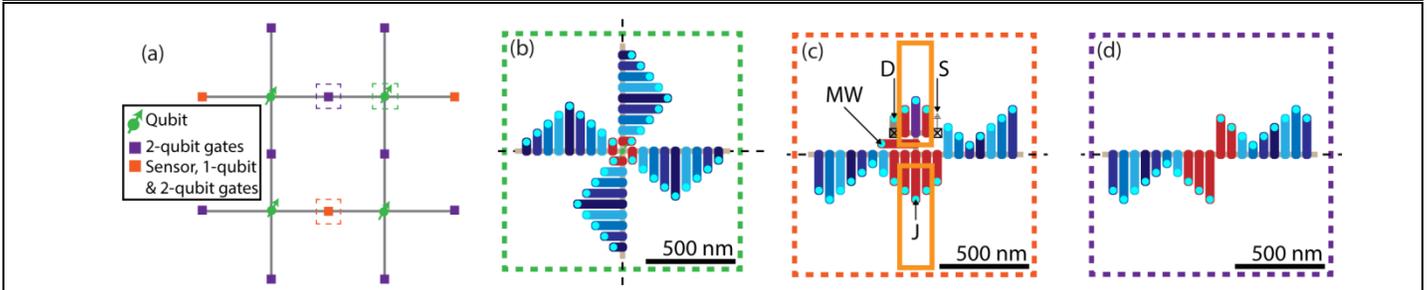

Fig. 3 (a) Schematic of a unit cell containing four spin qubits (green), operation regions (purple/orange), connected via shuttling channels (grey lines). (b) Qubit idling region. Four barrier gates (red) define the confinement potential and allow qubits into the shuttling channels (blue). Cyan circles represent vias. (c) Qubit operations region including control gates (red), sensing dot plunger (purple), source (S)/drain (D) ohmics (squares) and micromagnets (orange rectangles). (d) Two-qubit operation only regions.

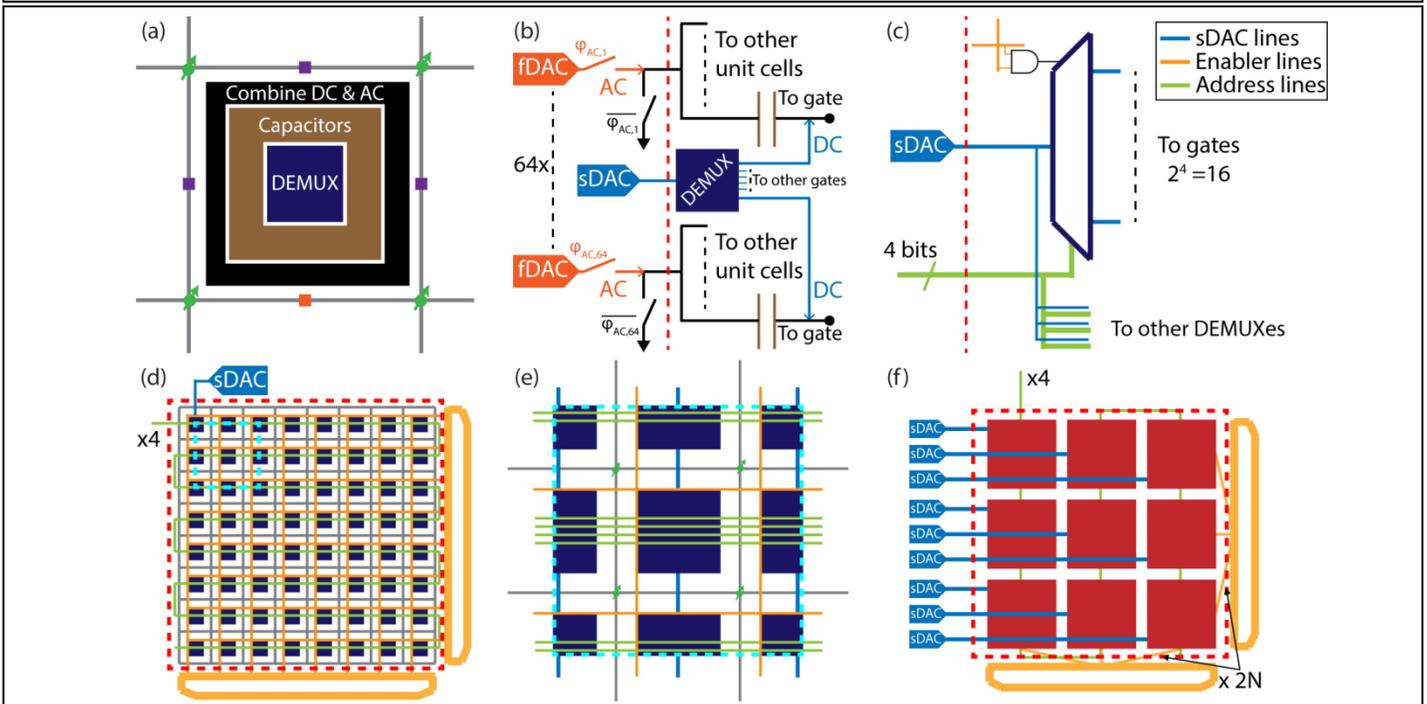

Fig. 4 (a) Schematic of a unit cell with locally integrated classical electronics. The color coding represents the same components in all the panels. (b) Circuit schematic of the components in (a), with the functionality described in the main text. fDAC (sDAC) are voltage sources for pulsed signals (DC biasing). Dashed red line denotes the quantum plane boundary in this and following panels. (c) Input/output schematic of the demultiplexer. (d) Schematic of a module. Demultiplexers are sequentially enabled by crossbar addressing controlled by multiplexers (orange blocks). (e) Zoom into the area surrounding a single unit cell in (d). (f) Schematic of the array of modules completing the quantum plane.